\documentclass[12pt]{article}
\usepackage{aasms}
\usepackage[dvips]{graphicx}
\usepackage{rotating}
\usepackage{lscape}
\usepackage{doublespace}

\begin{document}

\singlespace

\title{Correlations Between Spectral Properties and Spin-Down Rate 
in Soft Gamma--Ray Repeaters and Anomalous X--ray Pulsars}

\author{D. Marsden\footnote[1]{NAS/NRC Research Associate} and 
N. E. White}
\affil{NASA/Goddard Space Flight Center, Code 662, Greenbelt, MD 
20771}

\begin{abstract}

Anomalous x--ray pulsars (AXPs) and soft gamma--ray repeaters (SGRs) 
are x--ray sources with unusual properties distinguishing them from 
both rotation--powered and most accretion--powered pulsars. Using 
archival {\it ASCA} data over the energy range $0.5-10.0$ keV, 
we have studied the spectra of the persistent emission from these 
sources and their variation with spin-down rate. Using a single power 
law spectral model, we find that the overall hardness of the spectra 
increase with increasing spin-down rate, and therefore the spectral 
and spin-down mechanism are inextricably linked in these objects. 
In terms of the two-component blackbody plus power law spectral 
models, this correlation is seen as an increasing hardness of the 
high energy component with increasing spin-down rate, with the 
temperature of the low energy blackbody component remaining 
essentially constant. Also for the two component spectral model: 
the ratio of the $2-10$ keV power law and bolometric blackbody 
luminosities gradually increases with the spin-down rate. We 
discuss these results in terms of the various theoretical models 
for SGRs and AXPs.

\end{abstract}


\section{Introduction}
\label{intro}

Soft gamma--ray repeaters (SGRs) and anomalous x--ray pulsars (AXPs) 
are a group of radio-quiet soft x--ray pulsars with spin periods in 
the range $5-12$ s, rapid spin-down rates with no observed intervals 
of spin-up, and x--ray luminosities $L_{x}\sim 10^{35}$ ergs s$^{-1}$.  
Most of the SGRs and AXPs are associated with supernova remnants (e.g., 
\cite{marsden00} and references therein), implying that they are young 
neutron stars. The lack of Doppler shifts associated with binary orbital 
motion (\cite{mer98}; \cite{kaspi99}) imply that SGRs and AXPs are not 
members of typical high mass binary systems, although systems with stellar 
companions of $<1M_{\odot}$ and Thorne--\.{Z}ytkov systems (\cite{ghosh97}) 
are not constrained in most cases. Among the differences between SGRs and 
AXPs is the fact that SGRs occasionally emit multiple bursts of gamma--rays 
which have unique properties distinguishing them from other bursting sources 
(e.g., \cite{hurley2000a} for a recent review), and the AXPs appear to have 
softer x--ray spectra than the persistent SGR emission (\cite{stella98}; 
\cite{sonobe94}). 

The theoretical models describing SGRs and AXPs can be divided into 
two categories based on the enegy source powering the x--ray emission. 
The magnetar model (\cite{duncan92}) postulates that the x--ray emission 
and rapid spin-down of SGRs and AXPs are due to an unusually strong 
magnetic field ($B>10^{14}$ G), which may decay on a $10^{3}-10^{4}$ 
yr timescale and power the x--ray emission. In accretion-based models 
for SGRs and AXPs (Alpar 2000; Chatterjee, Hernquist \& Narayan 2000), 
SGRs and AXPs are normal neutron stars, and their spin-down and x--ray 
luminosities are due to the accretion of material from a fossil disk 
formed from supernova ejecta. Here we focus on the spectral properties 
of the persistent emission from these sources, and how they relate to 
their spin-down rates. A more extensive analysis of the temporal and 
spectral properties of the SGRs and AXPs sample will be presented 
elsewhere. 

\section{Data Analysis \& Results}
\label{obs}

{\it ASCA} consists of $4$ co-aligned grazing incidence x--ray 
telescopes producing an angular resolution of $\sim 1'$ over the 
energy range $0.5-10.0$ keV (\cite{tanaka94}). We only used data from 
the Gas Imaging Spectrometers (GISs; \cite{ohashi96}), because 
not all of the sources had usable Solid State Imaging Spectrometer 
data. The GISs are gas proportional scintillator detectors having 
moderate energy resolution ($\Delta E/E\sim 8\%$ at $6$ keV) and 
an effective field of view of $40'$. The data were screened using 
standard screening criteria and extracted using the XSELECT v2.0. 
The {\it ASCA} observations of all the SGRs and AXPs with known 
spin periods and period derivatives are listed in Table 1. Except 
for three sources, the on-source spectra were taken from circular 
extraction regions of radius $6'$ centered on each source, and 
background was accumulated from adjacent source--free regions. 
The AXP J1709-40 was $>10'$ off-axis and therefore highly distorted 
by the {\it ASCA} optics. For this source, on-source data were 
extracted from an elliptical extraction region elongated in 
an azimuthal direction around the optical axis. The AXPs 2259+59 
and 1841--05 are situated amongst bright emission from their 
associated supernova remnants CTB 109 (\cite{rho97}) and Kes 73 
(\cite{gotthelf97}; GV97), respectively. CTB 109 subtends almost 
half a degree and can be resolved by {\it ASCA}; we therefore 
extracted background data from an annular region $6'-7'$ from the 
AXP. Kes 73 is only $4'$ in diameter and too small to be completely 
resolved by {\it ASCA}, so we followed GV97 and used background from 
a nearby {\it ASCA} observation of the Galactic Ridge, and modeled 
the SNR emission using a $\sim 0.6$ keV thermal bremsstrahlung with 
Gaussian emission lines from Ne, Mg, Si, and S at $\sim 0.9$, $1.4$, 
$1.8$, and $2.4$ respectively. 

Before extraction for spectral analysis, all the {\it ASCA} data were 
searched for SGR-like bursts by binning the the data on a $0.1$ 
timescale and searching for bursts by eye. Bursts were only found 
in the September 1998 SGR 1900+14 data (see e.g. \cite{murakami99}), 
and only time intervals without bursts from this data set were included 
in our analysis. All spectra were extracted from the event data using 
XSELECT, rebinned to $256$ channels, and then grouped into bins 
containing at least $20$ counts per bin to facilitate the proper 
fitting of the spectra using chi-squared. Standard GIS response 
matrices were used for the spectral fitting, and ancillary response 
files were generated using the {\it ASCA} tool ASCAARF. The $0.5-10.0$ 
keV phase-averaged spectra of the SGRs and AXPs were fit to single power 
law and blackbody plus power law spectral models (modified by interstellar 
absorption) using XSPEC v11.0.1. The spectral parameters were averaged 
for sources with more than one observation. 

The pulse period $P$ determined for each object, and the assumed value of 
the period derivative $\dot{P}$, are listed in Table $1$. Details of the 
pulsar timing analysis will be presented elsewhere, and the results given 
here are not sensitive to the exact values of the pulsar periods. Figure 1 
shows the mean value of the single power law photon index for each object 
plotted as a function of the spin-down rate $|\dot{\Omega}|=2\pi\dot{P}/P^{2}$, 
which is proportional to the spin-down torque. Generally the AXP spectra were 
not well-fit by the single power law model, and require the addition of a 
low energy spectral component. The single power law model did fit the SGR 
spectra well in general, but a notable exception is the April 1998 SGR 
1900+14 data, which requires the two component model to adequately fit 
the spectrum (Woods et al. 1999). In spite of these uncertainties, the 
single power law photon index provides a good measure of the overall 
hardness of the spectrum, and Figure 1 indicates a distinct hardening 
of the SGR and AXP phase-averaged spectra with increasing spin-down rate. 
To investigate the spectral/spin-down rate correlation further, we have 
plotted the spectral parameters of the two component blackbody plus power 
law spectral model versus $|\dot{\Omega}|$ in Figure $2$. Shown are the 
blackbody temperature, power law photon index, and $L_{PL}/L_{BB}$, which 
is defined as the ratio between the $2-10$ keV power law and bolometric 
blackbody luminosities. The bolometric blackbody luminosity was calculated 
using $L_{BB}=A\sigma T^{4}$, where $A$ is the emitting area, $T$ is the 
blackbody temperature in Kelvins, and $\sigma$ is the Stefan-Boltzman 
constant. The blackbody emitting area is obtained from the fitted 
normalization and an assumed distance, but the ratio $L_{PL}/L_{BB}$ 
is independent of the distance. The blackbody temperatures are 
approximately constant, but with considerable scatter, as a function 
of $|\dot{\Omega}|$, but both the power law hardness and $L_{PL}/L_{BB}$ 
increase with increasing spin-down torque.

\section{Discussion}
\label{results}

We have found correlations between the spectra and spin-down rates of SGRs 
and AXPs. In terms of the power law plus blackbody spectral model, the 
power law component hardens and the ratio between the $2-10$ keV power 
law luminosity and the blackbody luminosity ($L_{PL}/L_{BB}$) increases 
gradually as the spin-down rate increases. These observed correlations 
can be used to constrain the models for SGRs and AXPs. With respect to 
the accretion model, the spectra of known accretion-powered pulsars in 
binary systems (XRBs) are generally harder than the SGR and AXP spectra 
over the energy range $0.5-10.0$ keV (\cite{whiteXX}), and there is no 
known correlation between the spectral parameters and the spin-down rates 
of XRBs. If SGRs and AXPs are accretion-powered, then they clearly must 
have different physical parameters -- such as accretion rate, 
magnetic field strength, and accretion geometry  -- than XRBs. 
Evolutionary scenarios of isolated neutron stars with supernova 
fallback disks (\cite{chatterjee00b}) favor $B\sim 10^{13}$ G for 
the the SGRs and AXPs, which are stronger than the typical fields of 
XRBs (Makashima et al. 1999, and references therein) and radio pulsars 
(e.g., \cite{taylor93}). The magnetic field alone cannot explain both the 
spectral softness and the spectral/spin-down rate correlation in the SGRs 
and AXPs, however, because the spectra and the spin-down rates of XRBs 
depend on both the magnetic field $B$ {\it and} the accretion rate 
$\dot{M}$. The spectra of accreting neutron stars have been calculated 
(e.g., \cite{bottcher00}) for stars with $B\sim 10^{12}$ G, but it is 
unclear if these models can produce the correct spectral shape 
for the values of $B$ and $\dot{m}$ necessary to produce the rapid 
spin-down of SGRs and AXPs. These spectral models assume isotropic 
emission of seed photons from close to the neutron star surface and 
ignore the effects of beaming, which become significant for $B>10^{12}$G. 
Since beaming reduces the Comptonization at the magnetospheric boundary 
layer where the hard emission would originate (M. B\"{o}ttcher, private 
communication), more complete calculations are needed before the accretion 
models for SGRs and AXPs can be tested on the basis of their x--ray spectra.  

In the magnetar model (\cite{thompson96}), both thermal and non-thermal 
x--ray emission is expected from SGRs and AXPs, with the non-thermal 
power-law emission originating from a hydromagnetic wind of particles 
in the magnetosphere accelerated by Alfv\`{e}n waves generated by the 
decaying magnetic field. The spin-down of magnetars is due to a 
combination of standard magnetic dipole radiation torque and torque 
from the wind (\cite{harding99}). Both of these torques increase 
strongly with magnetic field strength, and therefore the hardening 
of the power law spectral component with $|\dot{\Omega}|$ implies a 
similar hardening of the underlying Alfv\`{e}n wave spectrum with 
increasing $B$ if the SGRs and AXPs are magnetars. Unfortunately, we 
are unaware of any calculations of the non-thermal spectra of magnetars 
to compare with the data. The {\it luminosity} of the wind emission, 
however, should increase with magnetic field strength for a given wave 
amplitude (e.g., $L_{A}\propto B^{2}$; \cite{thompson98}). This is 
qualitatively consistent with the increase of the $L_{PL}/L_{BB}$ 
ratio with $|\dot{\Omega}|$, as there is no evidence that the thermal 
component of the SGRs and AXPs varies systematically with $|\dot{\Omega}|$. 

A strong constraint on the magnetar model is provided by $L_{PL}/L_{BB}$, 
because the blackbody luminosity is expected to be a much stronger function 
of the magnetic field strength than the power law luminosity (e.g., $L_{BB}
\propto B^{4.4}$; \cite{thompson96}). Assuming the power law luminosity $L_{PL}
\propto L_{A}$, $L_{PL}/L_{BB}$ should be a {\it decreasing} function of 
$B$ and the spin-down rate. The opposite trend is observed in Figure $2$, 
however, as $L_{PL}/L_{BB}$ increases with $|\dot{\Omega}|$ and the magnetic 
field strength. For consistency with the magnetar model, the power law 
component of the SGRs and AXPs must extend to energies much lower than 
the arbitrary $2$ keV cutoff energy used to compute $L_{pl}$. If the power 
law in the two component spectral model is extrapolated to $\sim 50$ 
eV, for example, then $L_{PL}/L_{BB}$ becomes a decreasing function of 
$|\dot{\Omega}|$ -- in accordance with the magnetar model\footnote{This 
ignores the dependence of the spin-down rate on $\Omega$, but since the 
sources have identical spin periods to within a factor of $\sim 2$ this 
would be an insignificant correction}. This implies that the SGR and AXP 
spectra must extend to the far-UV if they are magnetars. The spectral 
energy distributions of the sources must then break downward sharply 
for consistency with the optical flux measurements and upper limits 
(\cite{hulleman00}; \cite{kaplan00}). Observations of spectral breaks 
in the non-thermal persistent emission in the far-UV would be important 
evidence in support of the magnetar model, and additional multiwavelength 
observations of SGRs and AXPs are needed to constrain both the magnetar 
and accretion models for these sources.  

\section{Summary}
\label{summary}

We have analyzed the spectra of SGRs and AXPs over the energy range 
$0.5-10.0$ keV with {\it ASCA}, and have identified several interesting 
correlations between the spectra and the spin-down rates of these objects. 
For the single power law spectral model, the hardness of the spectra 
increase with increasing spin-down rate. For the two component blackbody 
plus power law spectral model, the temperature of the blackbody shows no 
evidence of a systematic variation with spin-down rate, but the hardness 
of the high energy power law component and the ratio of the $2-10$ keV 
power law luminosity to the blackbody luminosity both increase with increasing 
spin-down rate. These observations can be used to constrain the magnetar 
and accretion models for SGRs and AXPs. The increasing power law emission 
with spin-down rate is consistent with the magnetar model, but the emission 
from the power law must extend into the far-UV band for this model to be 
consistent with the x--ray data. In the context of the fossil disk model, 
the softness of the SGR and AXP spectra argue against accretion by analogy 
with the spectra of known neutron star accretors, but it is possible that 
high field accreting neutron stars with $B\sim 10^{13}$ may still be 
consistent with the data. Calculations of the x--ray spectra of both 
magnetars and accretion--powered neutron stars with $10^{13}$ G fields 
are needed to more adequately evaluate these models.
  
\acknowledgements

We thank Alice Harding and Markus B\"{o}ttcher for discussions. This research 
has made use of data obtained from the High Energy Astrophysics Science 
Archive Research Center (HEASARC), provided by NASA's Goddard Space Flight 
Center, and was performed while one of the authors (DM) held a National 
Research Council-GSFC Research Associateship. 

{}

\newpage
 
\begin{figure}
\plotone{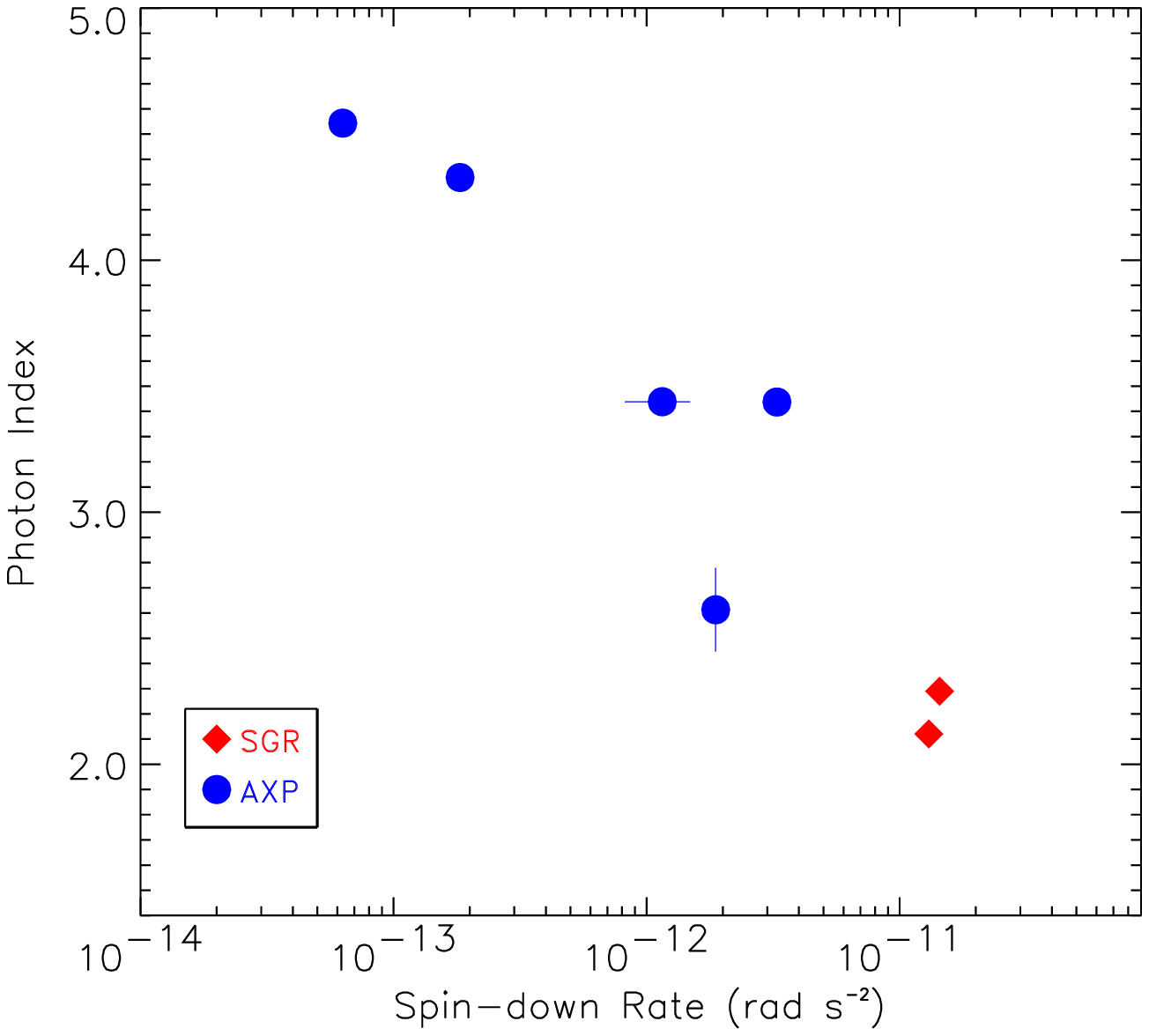}
\caption{~The variation of the single power law photon index versus 
spin-down rate $|\dot{\Omega}|$ for each SGR and AXP, where the results 
for objects with more than one observation have been averaged. The photon 
index decreases (spectral hardness increases) with increasing spin-down 
rate.}
\end{figure}

\begin{figure}
\centerline{\includegraphics[height=7in,keepaspectratio]
{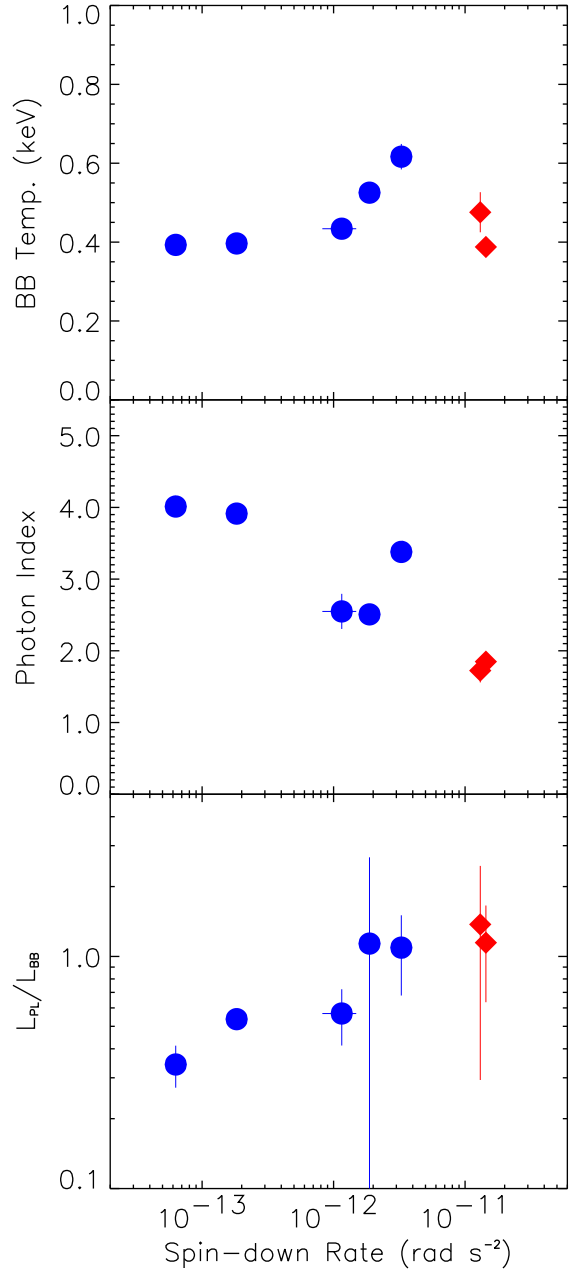}}
\caption{~Same as Figure 1, except for the blackbody plus power law 
spectral model. $L_{PL}/L_{BB}$ is the ratio between the $2.0-10.0$ 
keV power law luminosity and the bolometric blackbody luminosity.}
\end{figure}

\clearpage

\begin{planotable}{lccccr}
\tablewidth{0pt}
\tablecaption{SGR \& AXP Timing Parameters}
\tablehead{\colhead{Object} & \colhead{Start Date} & 
\colhead{$P$\tablenotemark{a}} & \colhead{$\dot{P}$\tablenotemark{b}} 
& \colhead{Refs.} \\ { } & (mm/dd/yy) & (s) &  ($10^{-12}$ s s$^{-1}$) 
& { }}
\startdata
SGR 1900$+$14 & 04/30/98 & $5.158971(7)$ & $61.0\pm 1.5 $ & 1,12\nl
SGR 1900$+$14 & 09/16/98 & $5.16025(2)$ & $61.0\pm 1.5$ & 2,12\nl
SGR 1806--20 & 10/10/93 & $7.468514(3)$ & $115.7\pm 0.2$ & 3,13\nl
SGR 1806--20 & 10/16/95 & $7.46445(3)$ & $115.7\pm$ 0.2& 3,13\nl
AXP 1048--59 & 03/03/94 & $6.446646(1)$ & $32.9\pm 0.3$ & 4,5\nl
AXP 1048--59 & 07/26/98 & $6.45082(1)$ & $16.7\pm 0.2$ & 5\nl
AXP 1841--05 & 10/11/93 & $11.76668(6)$ & $41.3\pm 0.1$ & 6,7\nl
AXP 1841--05 & 03/27/98 & $11.77243(7)$ & $41.3\pm 0.1$ & 7\nl
AXP 2259$+$59 & 05/30/93 & $6.97884(2)$ & $0.4883\pm 0.0003$ & 8,13\nl
AXP 2259$+$59 & 08/11/95 & $6.9788793(8)$ & $0.4883\pm 0.0003$ & 13\nl
AXP 0142$+$62 & 09/18/94 & $8.68794(7)$ & $2.2\pm 0.2$ & 9\nl 
AXP 0142$+$62 & 08/21/98 & $8.68828(4)$ & $2.2\pm 0.2$ & 5\nl
AXP 1709--40 & 09/03/96 & $10.99758(6)$ & $22\pm 6$ & 10,14\cr
\tablenotetext{a}{Measured period ($1\sigma$ error in last digit)}
\tablenotetext{b}{Assumed period derivative (from Refs.)}
\tablerefs{(1) \cite{hurley99a}; (2) \cite{murakami99}; (3) 
\cite{sonobe94}; (4) \cite{corbet97}; (5) \cite{paul00}; 
(6) \cite{gotthelf97}; (7) \cite{gotthelf99}; (8) \cite{corbet95}; 
(9) \cite{white96} ; (10) \cite{sugizaki97}; (11) \cite{woods99b}; 
(12) \cite{woods00}; (13) \cite{kaspi99}; (14) \cite{israel99}}
\end{planotable}

\end{document}